\documentclass[aps,prl,twocolumn,superscriptaddress,nofootinbib]{revtex4-2}
\usepackage{graphicx}% Include figure files
\usepackage{dcolumn}% Align table columns on decimal point
\usepackage{bm}% bold math
\usepackage{xcolor} % colored text
\usepackage{comment}
\usepackage{hyperref}
\usepackage{cancel}
\usepackage{amsmath} 
\usepackage{amssymb} 
\usepackage{amsfonts} 
\usepackage{slashed}
\usepackage{multirow}
\usepackage{amsfonts} 
\usepackage{mathtools}
\usepackage{ulem}
\newcommand{\CP}{{CP}}

\newcommand{\CPV}{{\cancel{\rm CP}}}
\newcommand{\bea}{\begin{eqnarray}}
\newcommand{\eea}{\end{eqnarray}}
\newcommand{\be}{\begin{equation}}
\newcommand{\ee}{\end{equation}}

\newcommand{\ba}{\begin{array}}
\newcommand{\ea}{\end{array}}

\newcommand{\mcA}{{\mathcal A}}

\newcommand{\mcD}{{\mathcal D}}
\newcommand{\mcE}{{\mathcal E}}

\newcommand{\mcN}{{\mathcal N}}

\newcommand{\Tr}{\mathrm{Tr}}

\newcommand{\lp}{\left}
\newcommand{\rp}{\right}
\newcommand{\la}{\langle}
\newcommand{\ra}{\rangle}

\begin{document}

\title{Calculation of neutron electric dipole moment  from Lattice QCD}% Force line breaks with \\

\author{Thomas Blum}%
% \email{thomas.blum@uconn.edu}
\affiliation{Physics Department, University of Connecticut, Storrs, Connecticut 06269, USA
}
\author{Fangcheng He}
\email{fangchenghe123@gmail.com}
\affiliation{Center for Nuclear Theory, Department of Physics and Astronomy, Stony Brook University, Stony Brook, New York 11794-3800, USA}
\affiliation{Department of Physics, New Mexico State University, Las Cruces, NM 88003, USA}
\affiliation{Nuclear Science Division, Lawrence Berkeley National Laboratory, Berkeley, CA 94720, USA}

\author{Taku Izubuchi}
\affiliation{Physics Department, Brookhaven National Laboratory, Upton, New York 11973, USA
}
\affiliation{RIKEN-BNL Research Center, Brookhaven National Laboratory, Upton, NY 11973, USA}

\author{Luchang Jin}
\affiliation{Physics Department, University of Connecticut, Storrs, Connecticut 06269, USA
}

\author{Hiroshi Ohki}
\affiliation{Department of Physics, Nara Women's University, Nara 630-8506, Japan}
\author{Sergey Syritsyn}
\affiliation{Center for Nuclear Theory, Department of Physics and Astronomy, Stony Brook University, Stony Brook, New York 11794-3800, USA}

\date{\today}% It is always \today, today,
                    
%%%%%%%%%%%%%%%%%%%%%%%%%%%%%%%%%%%%%%%%%%%%%%%%%%%%%%%%%%%%%%%%%%%%%%%%%%%%%%%%
\begin{abstract}
Experimental constraints on the neutron electric dipole moment (nEDM) may imply strong-CP problem in QCD, or unnatural smallness of the QCD theta angle.
In this work, we present a novel determination of the neutron electric dipole moment (nEDM) $d_n$ sensitivity to theta term from nonperturbative QCD on a lattice with background electric field.
Using Feynman-Hellmann theorem, we compute nEDM from the matrix element of local topological charge density between nucleon ground states spatially polarized by an electric field.
These states have mixed spatial parity, and we construct them using variational analysis.
We obtain statistically significant signal for the theta induced nEDM from lattices with 2+1 dynamical domain wall fermions corresponding to pion masses of 340, 420, and 576 MeV and lattice spacing $a\approx 0.11~\text{fm}$.
After extrapolating to the physical point, we obtain $d_n=-0.0050(4)(8)\bar{\theta}$ e$\cdot$fm.
Comparison with the current experimental bound on nEDM implies constraint $|\bar{\theta}|\lesssim 10^{-11}$, which confirms existence of the strong-CP problem in QCD.
Our pioneering work demonstrates that neutron EDM can be reliably determined from the local density of topological charge with robust control of systematic effects, and can be directly extended to other CP-violating interactions.
\end{abstract}
\maketitle
%%%%%%%%%%%%%%%%%%%%%%%%%%%%%%%%%%%%%%%%%%%%%%%%%%%%%%%%%%%%%%%%%%%%%%%%%%%%%%%%

%%%%%%%%%%

%\end{description}


%\tableofcontents

% I
%%%%%%%%%%%%%%%%%%%%%%%%%%%%%%%%%%%%%%%%%%%%%%%%%%%%%%%%%%%%%%%%%%%%%%%%%%%%%%%%
%%%%%%%%%%%%%%%%%%%%%%%%%%%%%%%%%%%%%%%%%%%%%%%%%%%%%%%%%%%%%%%%%%%%%%%%%%%%%%%%
\section{Introduction}
Charge–parity (CP) symmetry violation is one of the required conditions for the Baryonic asymmetry of the Universe
(BAU)~\cite{Sakharov:1967dj}.
 However, the small CP violation ($\CPV$)  arising from weak interactions through the Cabibbo-Kobayashi-Maskawa (CKM)
mechanism~\cite{Kobayashi:1973fv}  is  insufficient to explain the observed baryon
asymmetry~\cite{Sakharov:1967dj,Gavela:1993ts,Huet:1994jb}.
Additional $\CPV$ interactions  may arise either from strong interactions in the form of the hypothetical QCD $\Theta$-term, 
or outside the Standard Model in the form of effective interactions such as
quark chromo–electric interaction, Weinberg operator~\cite{Weinberg:1989dx}, four-quark operators, and other higher-dimensional operators.

Searches for intrinsic electric dipole moments (EDMs) of the nucleons, nuclei, and atoms are one of the most sensitive methods to probe CP violation.
The current best direct experimental bound of the neutron EDM is $d_n \sim
10^{-26}$ e$\cdot$cm~\cite{Abel:2020gbr}, and the indirect bounds $d_n<1.6 \times 10^{-26}$ e$\cdot$cm and $d_p<2.0\times
10^{-25}$ e$\cdot$cm are obtained from the experimental limit  on the $^{199}$Hg EDM~\cite{Graner:2016ses}.
These bounds are $(5-6)$ orders of magnitude larger than the nEDM predicted from weak
interaction~\cite{Khriplovich:1981ca,Czarnecki:1997bu,Seng:2014lea}.
The sensitivity of the experimental measurement of nEDM is expected to improve by 
up two order of magnitude over the next 10-20 years to $d_n \approx 3 \times 10^{-28}$ e$\cdot$cm at 90\% confidence~\cite{Alarcon:2022ero,Ito2019}.

From a theoretical perspective, it is highly challenging to determine the contributions of strong interactions to the
neutron EDM (nEDM) due to the non-perturbative nature of QCD at low energy scales.
Phenomenological approaches, including QCD sum
rules~\cite{Pospelov:1999ha,Pospelov:2000bw,Lebedev:2004va,Pospelov:2005pr,Fuyuto:2012yf,Haisch:2019bml,Ema:2024vfn},
chiral effective
theory~\cite{Crewther:1979pi,Pich:1991fq,Cho:1992rv,Borasoy:2000pq,Hockings:2005cn,Narison:2008jp,Ottnad:2009jw,deVries:2010ah,Mereghetti:2010kp},
and the instanton liquid model~\cite{Faccioli:2004jz,Liu:2025kuc}, have provided useful estimates of the relevant matrix
elements, but all of them are subject to sizable theoretical uncertainties.

Lattice QCD provides a first-principles framework for the calculation of CP violating matrix elements induced by the
strong interactions. 
The nEDM induced by QCD $\Theta$-term has been extensively studied in lattice
calculations~\cite{Shintani:2005xg,Berruto:2005hg,Shindler:2014oha,Guo:2015tla,Shindler:2015aqa,Alexandrou:2015spa,Shintani:2015vsx}.
Early Lattice studies employed an incorrect definition of the electric dipole form factor $F_3$, which contains the
spurious contribution from the Pauli form factor $F_2$~\cite{Abramczyk:2017oxr}.
Subsequent works have adopted the correct definition of the electric dipole form factors
~\cite{Dragos:2019oxn,Alexandrou:2020mds,Bhattacharya:2021lol,Liang:2023jfj}, and have also extended the analysis to the
quark chromo–EDM~\cite{Abramczyk:2017oxr,Kim:2018rce,Bhattacharya:2023qwf}; see Ref.~\cite{Liu:2024kqy} for a recent
review.
All these calculations extract nEDM from the forward limit of electric dipole form factor $F_3(Q^2)$, which is computed
from correlations of $\CP$-odd operators in the three-point correlation function of the quark vector current.
However, this correlator suffers from severe statistical noise due to the fluctuation of the global topological charge
operator.
As a result, it remains a challenge to obtain a statistically significant signal of  nEDM induced by $\Theta$-term at the
physical point~\cite{Alexandrou:2020mds,Bhattacharya:2021lol}. 
Even at  heavier quark masses, obtaining a nonzero signal requires a truncated-volume summation of the topological
charge operator~\cite{Dragos:2019oxn,Liang:2023jfj}, introducing additional systematic uncertainties.
Furthermore, the extrapolation of $F_3(Q^2)$ to the forward limit, $Q^2\to0$, also introduces some systematic
uncertainty.

An alternative strategy is the background field method, in which the nEDM is determined from the energy shift of the nucleon $\Delta E\propto d_n(\vec \Sigma\cdot\vec\mcE)$ in a uniform electric field $\vec\mcE$~\cite{Aoki:1989rx,Shintani:2006xr, Shintani:2008nt,Izubuchi:2007rmy,Abramczyk:2017oxr,Izubuchi:2020ngl,He:2023gwp}.
Building on this idea, we present a novel determination of the nEDM, in which
it is related to the matrix element of the local (single time-slice) topological charge evaluated in the nucleon ground state spatially-polarized by a uniform background electric field.
This ground state is a mixture of positive- and negative-parity states, which we construct variationally by solving a
generalized eigenvalue problem (GEVP).
This method offers the key advantage of suppressing excited state contamination 
while also controlling the spurious mixing between $F_2$ and $F_3$.
In addition, no $Q^2\to0$ extrapolation is needed.
It enables more reliable lattice determination of nEDM induced by $\CP$-violating topological charge and other interactions where statistical noise is worsened by global volume summation, such as Weinberg three-gluon and isosinglet four-quark effective operators.

%%%%%%%%%%%%%%%%%%%%%%%%%%%%%%%%%%%%%%%%%%%%%%%%%%%%%%%%%%%%%%%%%%%%%%%%%%%%%%
\section{Methodology}
The only dimension-four $\CPV$ operator generated by the strong interactions is the QCD $\Theta$-term.
The Euclidean QCD action is~\cite{Blum:2026hul}
\begin{equation}
\label{eq:actionME}
%\textbf{exp}
-S_{QCD,E}-i\bar{\theta}\int d^4x \frac{G^c_{\mu\nu}  {\widetilde G^c_{\mu\nu}} }{32\pi^2}, 
\end{equation}
where $\bar{\theta} = \theta_{QCD} + {\rm Arg\ Det}M_q$ is the physical theta-angle (including the phase of the quark mass matrix $M_q$).
$({\widetilde G^c_{\mu\nu}}) = \frac12\epsilon_{\mu\nu\rho\sigma} G^c_{\rho\sigma}$ is the dual gluon field strength, and
$S_{QCD,E}$ is the usual CP-even action.

To measure the nEDM, we first impose a uniform electric field that preserves the (anti)periodicity in time~\cite{Detmold:2009dx}.
For the electric field in the $z$ direction,
the usual $SU(3)$ gauge links $U_\mu(x)$ are modified by a $U(1)$ phase corresponding to background field, in the
spatial $z$ ($0\le z=x_3 < L_z$) and temporal ($0\le t=x_4 < L_t$) directions, 
\begin{equation}
\label{eq:gauge}
\begin{aligned}
U_3(x) &\rightarrow e^{-iQ_q\mcE_zL_zx_4} U_3(x)\,,
\quad x_3 = L_z-a\,,
\\
U_4(x) &\rightarrow e^{iQ_q\mcE_zx_3a} U_4(x)\,,
\quad 0\le x_4 < L_t\,,
\end{aligned}
\end{equation}
where $Q_q$ is the quark electric charge and $a$ is the lattice spacing.
The spatial gauge link in the $z$ direction is modified 
only at the boundary $x_3=L_z-a$ to ensure periodicity of the $U(1)$ component of plaquettes.
To maintain gauge invariance on the torus, the uniform background field must satisfy the ’t Hooft quantization
condition~\cite{tHooft:1979rtg,tHooft:1981nnx}, 
$|Q_d|\mcE_z L_zL_t=2\pi n_z$ with integer $n_z$.
We choose four values for $\mcE_z$,
\begin{equation}
\label{eq:ezc}
\mcE_z=\frac{6\pi}{L_z L_t}n_{z}\,, \quad n_{z}=\pm1,\pm2.  
\end{equation}

The electric field in Euclidean space $\mcE$ corresponds to an analytic continuation from Minkowski space, $\mcE_M\rightarrow -i\mcE$. 
To linear order in $\mcE_z$ and $\bar\theta$, a real-valued EDM induces an imaginary energy shift of the neutron,
\begin{equation}
\label{eq:eneshift}
E_N = m_N + i \bar{\theta} (d_n/\bar{\theta})\,(\vec\Sigma \cdot \vec\mcE)\,,
\end{equation}
where $\Sigma_z=-i\gamma_x\gamma_y$ is the nucleon spin vector.  
We treat the $\CP$-odd interaction as infinitesimal and extract the induced nEDM from this first-order expansion.

The complete two-point nucleon correlation function 
in Euclidean space is given by
\begin{equation}
C_{2pt,\mcE,\bar\theta} =\la O(t_f) \, O^\dag(0) \, e^{ -i\bar{\theta} Q_{top} -i\int d^4x \mcA^\mu(x)J^\mu_{EM} }\ra_{S_{QCD}},
\end{equation}
where the electromagnetic vector potential corresponding to the electric field~(\ref{eq:gauge}) can be chosen as $\mcA^\mu=(0,0,0,z\mcE_z)$. This background field induces an anti-hermitian electric dipole interaction in the Hamiltonian,
\begin{equation}\label{eq:Dz}
\Delta H = i\mcE_z \mcD_z\,, 
\end{equation}
where $\mcD_z = \int d^3r \, z \,\sum_q Q_q q^\dag q\,$ is an electric dipole operator. This term is CP-even but explicitly
breaks parity symmetry.

According to the Feynman-Hellmann theorem~\cite{Feynman:1939zza}, the energy shift of the spin-up nucleon state can be related to
the matrix element of the local topological charge operator, $q_{top}(t)= \int d^3x \frac{G^c_{\mu\nu}  {\widetilde G^c_{\mu\nu}}(t,\vec x) }{32\pi^2}$, 
which is defined as the spatial sum of the topological charge density at a single time slice. The Feynman–Hellmann relation then gives
\begin{equation}
\label{eq:localq}
\frac{\partial E^\uparrow_N}{\partial \bar{\theta}}\Big{|}_{\bar{\theta}=0} 
= i \la \mathcal{N}^{L,\uparrow}_0|q_{top}(0)|\mathcal{N}^{R,\uparrow}_0\ra\Big|_{\mcE_z}\,. 
\end{equation}
Here, $\la \mathcal{N}^{L,\uparrow}_0|$ and $|\mathcal{N}^{R,\uparrow}_0\ra$  denote the biorthogonal unit-normalized left
and right eigenstates of  the nucleon ground state in the presence of the Euclidean background electric
field. The distinction arises because the  dipole interaction Eq.~(\ref{eq:Dz}) is anti-Hermitian, rendering
the effective Hamiltonian non-Hermitian. 
The ground state can be obtained as the first-order non-Hermitian perturbation,
\begin{equation}
\label{eq:Npm}
\begin{aligned}
|\mcN^R_0\ra &= |N_0\ra
+ i\mcE_z \sum_{i} \frac{\la N^*_i|\mcD_z|N_0\ra}{m_{N_0} - m_{N^*_i}} |N^*_i\ra + O(\mcE_z^2) \,,\\
\la\mcN^L_0| &= \la N_0| 
+ i\mcE_z \sum_{i} \frac{\la N_0|\mcD_z|N^*_i\ra}{m_{N_0} - m_{N^*_i}} \la N^*_i| + O(\mcE_z^2)\,, 
\end{aligned}
\end{equation}
where $|N_i\ra$ and $|N^*_i\ra$, $i=0,1,\ldots$ are towers of positive- and negative-parity
states for $\vec\mcE=\bar\theta=0$, respectively.

Combining Eqs.~(\ref{eq:eneshift}) and (\ref{eq:localq}), the $\bar{\theta}$-induced nEDM can be determined from the matrix element of
the local topological charge operator,
\begin{equation}
\label{eqn:edm_bgem_plateau0}
d_n/\bar{\theta} = \frac{1}{\mcE_z} \la \mathcal{N}_0^{L,\uparrow} |q_{top}(0)| \mathcal{N}_0^{R,\uparrow}\ra_{\mcE_z}. 
\end{equation}
Using the eigenstates~(\ref{eq:Npm}), the nEDM~(\ref{eqn:edm_bgem_plateau0}) can be expressed as
\begin{equation}\label{eq:ednNNs}
d_n/\bar{\theta}
= 2\sum_i 
  \frac{ \mathrm{Im}\lp[\la N_0|q_{top}|N^*_i\ra \la N^*_i|\mcD_z |N_0\ra\rp]}
       { m_{N^*_i}-m_{N_0}}.
\end{equation}
Equation~(\ref{eq:ednNNs}) explicitly shows that
the nEDM arises from the interplay between the CP-even electric dipole operator and the CP-odd topological charge operator.
The above equation is consistent with the one given in~\cite{Baluni:1978rf,Baym:2016lyf}, 
except the Euclidean topological charge density is imaginary~\cite{Blum:2026hul}.

The matrix element defined in Eq.~(\ref{eqn:edm_bgem_plateau0}) can be extracted on the lattice from the ratio of a
lattice three- and two-point correlation functions,
\begin{equation}
\label{eqn:edm_bgem_plateau}
d_n/\bar{\theta} 
\approx \frac{1}{\mcE_z} 
  \frac{ \la O^{L,\uparrow}_{\mathcal{N}_0}(t_f) \, q_{top}(\tau) \,  O^{\dag R,\uparrow}_{\mathcal{N}_0}(0)\ra_{\mcE_z}}
       { \la O^{L,\uparrow}_{\mathcal{N}_0}(t_f) \, O^{\dag
R,\uparrow}_{\mathcal{N}_0}(0)\ra_{\mcE_z}}\Bigg|_{t_f,\tau,
t_f-\tau\rightarrow\infty}\,,
\end{equation}
where $O^L_{\mathcal{N}_0}(0)$ and $O^R_{\mathcal{N}_0}(0)$ are interpolating operators optimized to
project onto the ground nucleon state in the presence of a background field. 
These interpolating operators must include both parity components to reflect
the spatial parity breaking induced by the electric dipole interaction. Starting from a nucleon interpolating operator $O$, definite-parity components are constructed as $(1\pm\gamma_4)O$. The operators $O^L_{\mathcal{N}_0}$  and $O^R_{\mathcal{N}_0}$ are then obtained as linear combinations of these components by solving the generalized-eigenvalue problem (GEVP) for the following correlation matrix,
\begin{equation}\label{eq:2ptppbar}
C^{\psi\psi^\dag,\uparrow}_{2pt,\mcE}(t_f)
=\left(\begin{array}{cc} \la\psi^\uparrow_N(t_f) \psi^{\uparrow,\dag}_{N}(0)\ra_{\mcE_z}  & \la\psi^\uparrow_N (t_f)\psi^{\uparrow,\dag}_{N^*}(0)\ra_{\mcE_z}  \\
\la\psi^\uparrow_{N^*}(t_f) \psi^{\uparrow,\dag}_{N}(0)\ra_{\mcE_z} &  \la\psi^\uparrow_{N^*}(t_f) \psi^{\uparrow,\dag}_{N^*}(0)\ra_{\mcE_z}\end{array}\right),
\end{equation}
where $\psi^\uparrow_N,\psi^\uparrow_{N^*}=\frac14(1\pm\gamma_4)(1+\Sigma_z)O$ 
denote the spin-up, positive- and negative-parity components of the nucleon interpolating operator $O$.
We retain only the spin up components for simplicity (the analysis for the spin down state is similar). The correlation matrix in Eq.~(\ref{eq:2ptppbar}) is non-hermitian
due to the non-hermitian Hamiltonian. 
In this case, one can define the left  and right generalized eigenvectors through 
\bea\label{eq:nonHGEVP}
v^\dagger_{L,n}C^{\psi\psi^\dag,\uparrow}_{2pt,\mcE}(t_f)v_{R,n} &=&\lambda_n v^\dagger_{L,n}  C^{\psi\psi^\dag,\uparrow}_{2pt,\mcE}(t_0)v_{R,n}, 
\eea
here $v_{L,n}$, $v_{R,n}$ and $\lambda_n$ represent the left eigenvector, 
right eigenvector, and generalized eigenvalue corresponding to the $n$-th state. The eigenvalue $\lambda_n$ is real owing to the pseudo-Hermiticity of $C^{\psi\psi^\dag,\uparrow}_{2pt,\mcE}(t_f)$~\cite{Blum:2026hul}. The ratio constructed using optimal  interpolating operators defined in Eq.~(\ref{eqn:edm_bgem_plateau}) 
can be obtained from the lowest-state eigenvectors,
\begin{equation}
\label{eqn:edm_bgem_GEVP1}
d_n^\theta(t_f,\tau)
=\frac{1}{\mcE_z} 
  \frac{ v^\dag_{L,0}C^{\psi\psi^\dag,\uparrow}_{3pt,\mcE}(t_f,\tau)v_{R,0}}       {v^\dag_{L,0}C^{\psi\psi^\dag,\uparrow}_{2pt,\mcE}(t_f)v_{R,0}}\xRightarrow{t_f,\tau,t_f-\tau\rightarrow\infty}d_n/\bar{\theta}.
\end{equation}
In the large time-separation limit, this ratio saturates the ground-state matrix element and yields the nEDM. The three point correlation matrix $C^{\psi\psi^\dag,\uparrow}_{3pt,\mcE}(t_f,\tau)$ is given by
\begin{equation}
\label{eq:3ptoriQtop}
\begin{aligned}
&C^{\psi\psi^\dag,\uparrow}_{3pt,\mcE}(t_f,\tau)
\\
&=\left(\begin{array}{cc} \la\psi^\uparrow_N(t_f) q_{top}(\tau)\psi^{\uparrow,\dag}_{N}(0)\ra_{\mcE_z}  & \la\psi^\uparrow_N (t_f)q_{top}(\tau)\psi^{\uparrow,\dag}_{N^*}(0)\ra_{\mcE_z}  \\
\la\psi^\uparrow_{N^*}(t_f)q_{top}(\tau)\psi^{\uparrow,\dag}_{N}(0)\ra_{\mcE_z} &  \la\psi^\uparrow_{N^*}(t_f) q_{top}(\tau)\psi^{\uparrow,\dag}_{N^*}(0)\ra_{\mcE_z}\end{array}\right).
\end{aligned}
\end{equation}

% III
%%%%%%%%%%%%%%%%%%%%%%%%%%%%%%%%%%%%%%%%%%%%%%%%%%%%%%%%%%%%%%%%%%%%%%%%%%%%%%%%
%%%%%%%%%%%%%%%%%%%%%%%%%%%%%%%%%%%%%%%%%%%%%%%%%%%%%%%%%%%%%%%%%%%%%%%%%%%%%%%%
\section{Numerical results}
In this study, we use three gauge ensembles generated by the RBC/UKQCD collaborations with 2+1 flavors of domain wall fermions (DWF) 
and the Iwasaki gauge action~\cite{Allton:2008pn}. The lattice volume is $(L/a)^3\times L_t/a=24^3\times64$, with 
inverse lattice spacing $a^{-1}=1.785$ GeV. 
The ensemble details are summarized in Tab.~\ref{Tab:ensembles}.
%--------------------------------------------------------------------
\begin{table}[t]
	\caption{\label{Tab:ensembles}
	The parameters of gauge ensembles. Here, $N_\text{cfg}$ is the number of configurations used in the calculation. For more details, see~\cite{Allton:2008pn}.}
	\begin{ruledtabular}
		\begin{tabular}{c c c c c }
			\text{ensemble} & $(L/a)^3 \times L_t/a$   &  $a$ (fm)   & $m_\pi$ (MeV) &   $N_\text{cfg}$\\
			\hline
			24I-005             & $24^3 \times 64$ & 0.1105(3)   & 340       &  1400\\
            \hline
            24I-010             & $24^3 \times 64$ & 0.1105(3)   & 420       & 1100  \\
	        \hline
            24I-020             & $24^3 \times 64$ & 0.1105(3)   & 576       & 523
		\end{tabular}
	\end{ruledtabular}
\end{table}

In this calculation, we employ the 
electro-quenched approximation, in which only the valence quarks interact with the background field. The topological charge is constructed from the gauge field evolved with the gradient flow~\cite{Luscher:2010iy,Luscher:2011bx}.
We consider four nucleon interpolating operators,
\begin{equation}
\label{eqn:nucleon_ops}
\begin{aligned}
O_1 &= \epsilon^{abc}(d^{aT}C\gamma_5u^b)d^c,
&\hspace{-0.1em}
O_2 &= -\epsilon^{abc}(d^{aT}Cu^b)\gamma_5d^c,\\
O_3 &= \epsilon^{abc}(d^{aT}C\gamma_4\gamma_5u^b)d^c,
&\hspace{-0.1em}
O_4 &= \epsilon^{abc}(d^{aT}C\gamma_4u^b)\gamma_5d^c .
\end{aligned}
\end{equation}

\begin{figure}[ht!] 
\centering
\includegraphics[width=.4\textwidth]{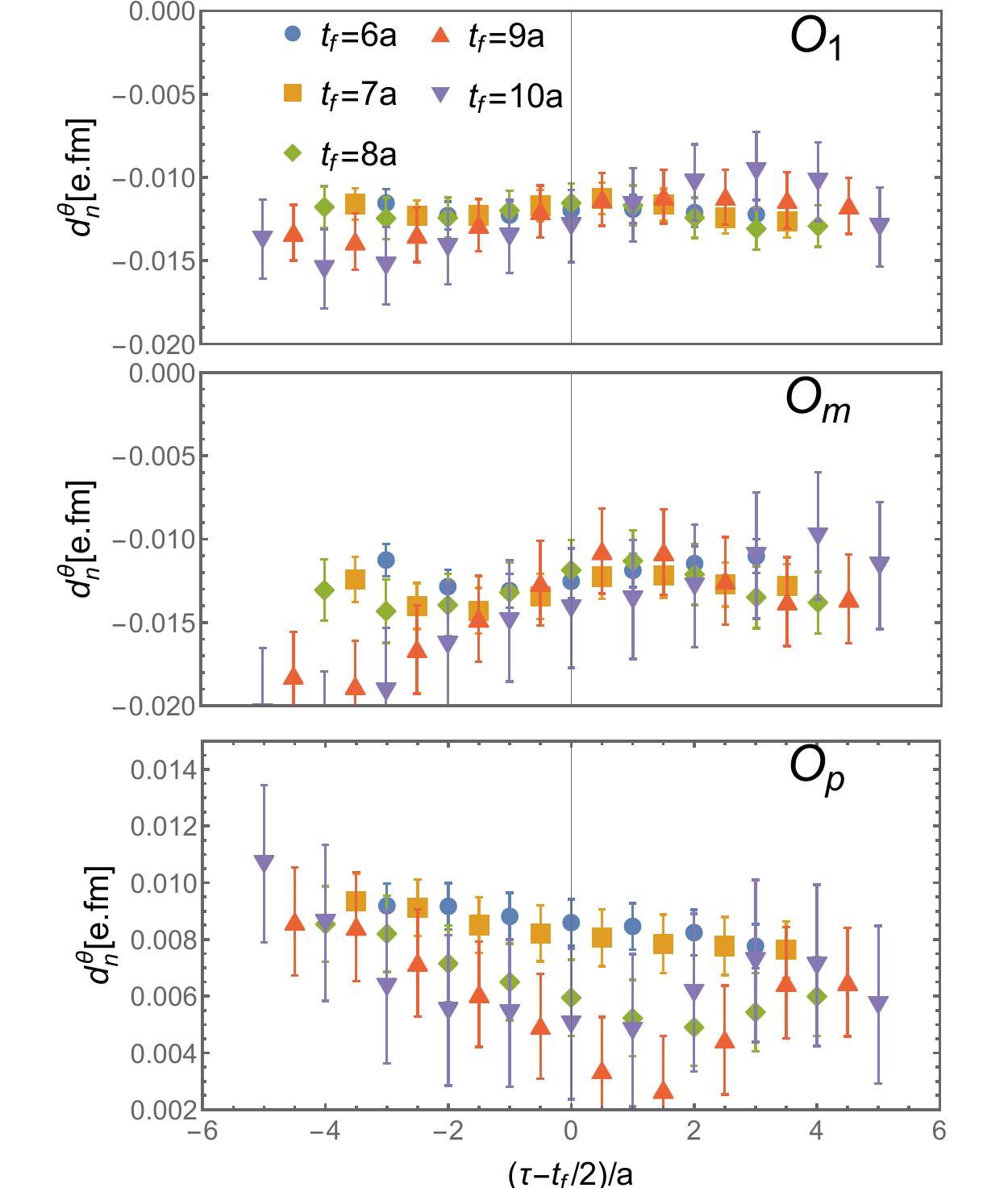}\\
\caption{$d_n^\theta(t_f,\tau)$ (Eq.~(\ref{eqn:edm_bgem_GEVP1})) with interpolating operators $O_1$, $O_m$, and $O_p$ on ensemble 24I-005 with electric field strength $|n_z|=1$.}
\label{fig:GEVPres_nz1}
\end{figure}

One can define $O_p=O_1+O_2$ and $O_m=O_1-O_2$, which are covariant under axial $U(1)$  transformation~\cite{Ema:2024vfn}.
In Figure~\ref{fig:GEVPres_nz1}, we present the numerical results for $d_n^\theta(t_f,\tau)$ with electric field strength $|n_z|=1$ obtained using interpolating operators $O_1$, $O_p$ and $O_m$. 
The results from $O_1$ and $O_m$ are consistent with each other,  whereas $O_p$ exhibits large excited state contamination.
As shown in the Supplemental Material~\cite{supp}, the operator $O_p$ has a relatively small overlap with a certain excited state. Consequently, the GEVP analysis does not effectively suppress the corresponding contamination to the ground state matrix element for $O_p$.

%--------------------------------------------------------------------

Excited state contamination can be further reduced by combining nucleon correlation functions 
from multiple interpolating operators in a GEVP~\cite{Blossier:2009kd,Hackl:2024whw}.
The corresponding two- and three-point correlation matrices are
\begin{equation}
\begin{aligned}
C_{\{2pt,\mcE\},ij}(t_f) &= \la O_i(t_f) O_j^\dag(0)\ra_{\mcE_z}, \\
C_{\{3pt,\mcE\},ij}(t_f,\tau) &=\la O_i(t_f) q_{top}(\tau)O_j^\dag(0)\ra_{\mcE_z}
\end{aligned}
\end{equation}
For each combination of $(O_i, O_j^\dag)$, the correlation function is a $2\times 2$ matrix of positive- and negative-parity projected operators similar to Eq.~(\ref{eq:2ptppbar}) and Eq.~(\ref{eq:3ptoriQtop}).
In the following, we present results obtained from a GEVP analysis 
using the full operator basis $\{O_1,O_2,O_3,O_4\}$ with the GEVP reference time fixed at $t_0=5a$. Results obtained with $t_0=4a$, $5a$, and $6a$ are consistent within statistical uncertainties. We also observe that our nEDM results are consistent over a wide range of gradient flow times for the topological charge~\cite{supp}.

\begin{figure}[ht!] % Fig. 7
\centering
\includegraphics[width=.4\textwidth]{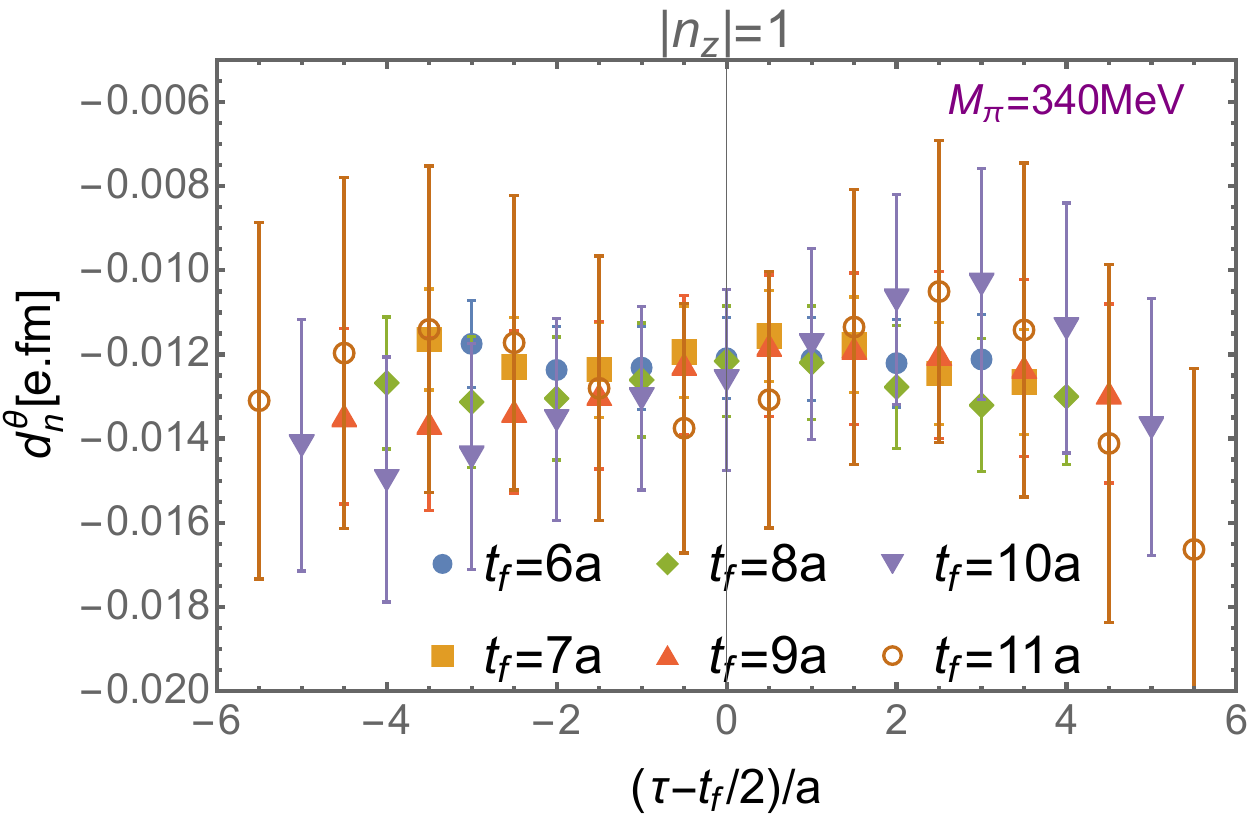}
\caption{The ratio $d_n^\theta(t_f,\tau)$, obtained from a GEVP analysis 
using the full operator basis $\{O_1,O_2,O_3,O_4\}$,  on ensemble 24I-005 with electric field strength $|n_z|=1$.}
\label{fig:multiGEVP_res_igf4}
\end{figure}

In Fig.~\ref{fig:multiGEVP_res_igf4}, we present the numerical results for $d_n^\theta$ with electric field strength $|n_z|=1$ on ensemble 24I-005. The source sink 
separation $t_f$ ranges from 6a to 11a. The mild dependence on $t_f$ indicates that excited state contamination is negligible. To extract the ground state matrix element, we perform a constant fit to the results at $t_f=8a,9a,10a$. 
 For each value of $t_f$, we select the three central points and carry out a correlated constant fit. 
 To estimate the systematic uncertainty associated with the fit range, we repeat the fit using the data points for $t_f=9a,10a$, and $11a$ and take difference between the two as an estimate of the systematic uncertainty. 
The analysis strategy on the other ensembles is similar.

%--------------------------------------------------------------------

\begin{figure}[ht!] % Fig. 7
\centering
\includegraphics[width=.4\textwidth]{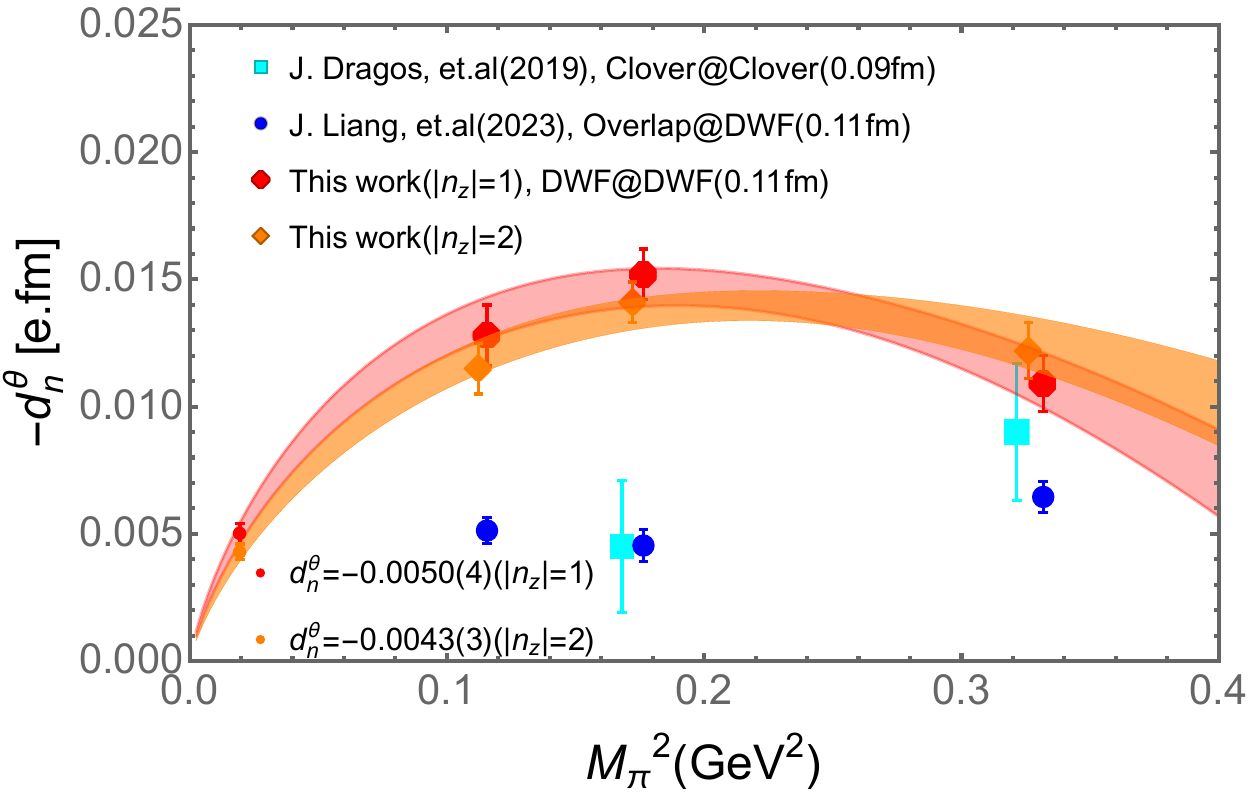}
\caption{Comparison of lattice nEDM results. 
The red and orange bands show our chiral extrapolation for two different electric field strengths. The leftmost points are from the fits evaluated at the physical pion mass, with their values indicated in the plot.}
\label{fig:EDM_compa}
\end{figure}

The nEDM results on three ensembles are shown in Fig.~\ref{fig:EDM_compa}. The data points with filled red circles and orange diamonds correspond to 
the electric field strengths $|n_z|=1$ and $|n_z|=2$, respectively. The consistency between them indicates that the applied electric fields are sufficiently small so that higher-order contributions can be neglected. For comparison, we also include results from~\cite{Dragos:2019oxn} (cyan squares) where clover fermions are used for both valence and sea quarks with a lattice spacing approximately 0.09 fm. The blue circles are from Ref. \cite{Liang:2023jfj}, which uses the same gauge field ensembles as in this work but employs overlap fermion valence quarks. At the heaviest pion mass, our results are roughly comparable with theirs. However, there is a significant difference for lighter pion masses. 

%--------------------------------------------------------------------

In the presence of the electric field, parity is no longer a good quantum number, and  
the physical hadron states are polarized mixtures of positive- and negative-parity states. 
Our GEVP analysis isolates these eigenstates and suppresses excited state contamination, 
ensuring consistency between results obtained with different nucleon operators. 
However, as we show in the Supplemental Material~\cite{supp} and Ref.~\cite{Blum:2026hul}, the results obtained using the conventional positive-parity projection $(1+\gamma_4)$  exhibit strong excited state contamination and depend noticeably on the nucleon interpolating operators used.
These residual excited state effects may explain the observed discrepancy between our results and those in previous studies of the $F_3$ form factor~\cite{Dragos:2019oxn,Liang:2023jfj}. Differences in lattice actions, lattice spacings, volumes may also contribute to the discrepancy.

Results obtained at different pion masses are extrapolated to the physical point using the ansatz~\cite{Crewther:1979pi, OConnell:2005mfp},
\begin{equation}
d_n/\bar\theta=c_0 m_\pi^2+c_1m_\pi^2\log\left(\frac{m_\pi^2}{m^{2}_{N,phy}}\right).
\end{equation}
The red and orange bands in Fig.~\ref{fig:EDM_compa} represent the extrapolations for the two electric field strengths. 
At the physical point, we obtain $d_n/\bar\theta=-0.0050(4)(4)$ for $|n_z|=1$ and $d_n/\bar\theta=-0.0043(3)(3)$ for $|n_z|=2$, where the first uncertainty is statistical and the second one arises from the choice of fit region. 
We quote the results of $|n_z|=1$ as our central value, as it is expected to have smaller higher-order corrections in $\mcE_z$.  The total systematic uncertainty is obtained by adding the uncertainty of the fit region and the difference between the two $|n_z|$ results in quadrature.
Our value for the nEDM at the physical point is then $d_n/\bar\theta=-0.0050(4)(8)$.

A summary of recent lattice results for the contribution of the QCD $\Theta$-term to the nEDM is presented in Table~\ref{tab:Mix}. 
The studies in ~\cite{Dragos:2019oxn,Liang:2023jfj} extract EDM from vector current form factors using a truncated topological charge sum to reduce statistical noise, however, the choice of the truncated region may introduce additional systematic uncertainty. 
The studies in Ref.~\cite{Alexandrou:2020mds,Bhattacharya:2021lol} are similar but at the physical point and employ global, full-4D volume topological charge, which yields no statistically significant nEDM signal.
In contrast, topological charge is sampled locally in our method, which avoids  global summation or space-time truncation and improves the statistical and systematic precision of nEDM determination.

%--------------------------------------------------------------------
\begin{table}[t]
    \centering
    \begin{tabular}{|c|c|} \hline
    &  Neutron EDM(e.fm)
  \\
  \hline
  Dragos et al(2019)\cite{Dragos:2019oxn}  &  $d_n=-0.00152(71)\bar{\theta}$    \\
  \hline
  Alexandrou et al(2020)\cite{Alexandrou:2020mds}  &  $|d_n|=0.0009(24)|\bar{\theta}|$  \\
  \hline  
  Bhattacharya et al (2021) \cite{Bhattacharya:2021lol} & $d_n=-0.028(18)(54)\bar{\theta}$  \\
  \hline 
  Liang et al (2023) \cite{Liang:2023jfj} & $d_n= -0.00148 (14) (31)\bar{\theta}$\\
  \hline
  This work & $d_n= -0.0050(4)(8)\bar{\theta}$ \\
 \hline 
    \end{tabular}
    \caption{Summary of lattice results for the QCD $\Theta$-term contribution to the  nEDM.}
    \label{tab:Mix}
\end{table}

% IV
%%%%%%%%%%%%%%%%%%%%%%%%%%%%%%%%%%%%%%%%%%%%%%%%%%%%%%%%%%%%%%%%%%%%%%%%%%%%%%%%
%%%%%%%%%%%%%%%%%%%%%%%%%%%%%%%%%%%%%%%%%%%%%%%%%%%%%%%%%%%%%%%%%%%%%%%%%%%%%%%%
\section{Conclusions}
In this work, we have calculated $\bar{\theta}$-induced nEDM using the novel method based on local topological charge density.
It reduces stochastic and avoids systematic uncertainties that may affect previous determinations based on global or truncated topological charge.
We have obtained a clear signal for $\bar{\theta}$-induced nEDM
$d_n=-0.0050(4)(8)\bar{\theta}$ e$\cdot$fm
by probing the topological charge density in spatially-polarized neutron states constructed variationally as mixtures of  positive- and negative-parity states, and extrapolating to the physical pion mass.
Combined with the current experimental bound on the neutron EDM, this value implies $|\bar{\theta}|\lesssim 10^{-11}$ and confirms the existence of the strong-CP problem.
To validate our method, we have varied the basis of nucleon interpolating operators and found that the results are consistent, as long as excited state contamination is properly controlled.

Our result may still be subject to systematic uncertainties arising from finite volume and discretization effects, as well as extrapolation to the physical pion mass from the lowest $m_\pi=340$ MeV point.
These uncertainties can be readily investigated by extending our method to lighter pion masses and including ensembles with smaller lattice spacing and larger volume.
Finally, our method can also be applied to study the effects of the Weinberg three-gluon operator, $\CPV$ four-quark operators, and other sources of $\CP$ violation.

%%%%%%%%%%%%%%%%%%%%%%%%%%%%%%%%%%%%%%%%%%%%%%%%%%%%%%%%%%%%%%%%%%%%%%%%%%%%%%
\section{Acknowledgements}
We thank Tanmoy Bhattacharya, Kaori Fuyuto, Rajan Gupta, Keh-Fei Liu, Jian Liang, Emanuele Mereghetti, Andrea Shindler for fruitful discussions. We thank Michael Abramczyk for valuable discussions and contributions during the initial stages of this project. FH and SS are supported by the National Science Foundation under award PHY-2412963. FH is also supported by the U.S. Department of Energy, Office of Science, Office of Nuclear Physics, under Grant No. DE-SC0013065. TB is supported by DOE grant DE-SC0010339. LJ is supported by DOE grant DE-SC0021147, DE-SC0010339, and DE-SC0026314. TI is supported by US DOE Contract DESC0012704(BNL) and the Scientific Discovery through Advanced Computing (SciDAC) program LAB 22-2580, and also Laboratory Directed Research and Development (LDRD No. 23 - 051) of BNL and RIKEN-BNL Research Center. (Any opinions,
findings, and conclusions or recommendations expressed in this material are those of the author(s) and do not
necessarily reflect the views of the National Science Foundation.)
HO is supported by the JSPS KAKENHI (Nos. 21K03554, 22H00138).

\bibliography{ref}% Produces the bibliography via BibTeX.
\clearpage

%%%%%%%%%%%%%%%%%%%%%%%%%%%%%%%%%%%%%%%%%%%%%%%%%%%%%%%%%%%%%%%%%%%%%%%%%%%%%%
%%%%%%%%%%%%%%%%%%%%%%%%%%%%%%%%%%%%%%%%%%%%%%%%%%%%%%%%%%%%%%%%%%%%%%%%%%%%%%
\onecolumngrid
\section*{Supplemental Material}
\subsection{Comparison of different interpolating operators}
In this section, we compare the nEDM results, effective masses, and biorthogonal weights obtained using different interpolating operators $O_1$, $O_p=O_1+O_2$ and  $O_m=O_1-O_2$. We also compare the nEDM results extracted from the positive-parity components of these interpolating operators using the conventional
positive-parity projection.
The operators $O_p$ and $O_m$ are covariant under a $U(1)$ axial rotation, whereas $O_1$ is not~\cite{Ema:2024vfn}.  
\subsubsection{nEDM results}
\begin{figure*}[ht!]
\centering
\includegraphics[width=.96\textwidth]{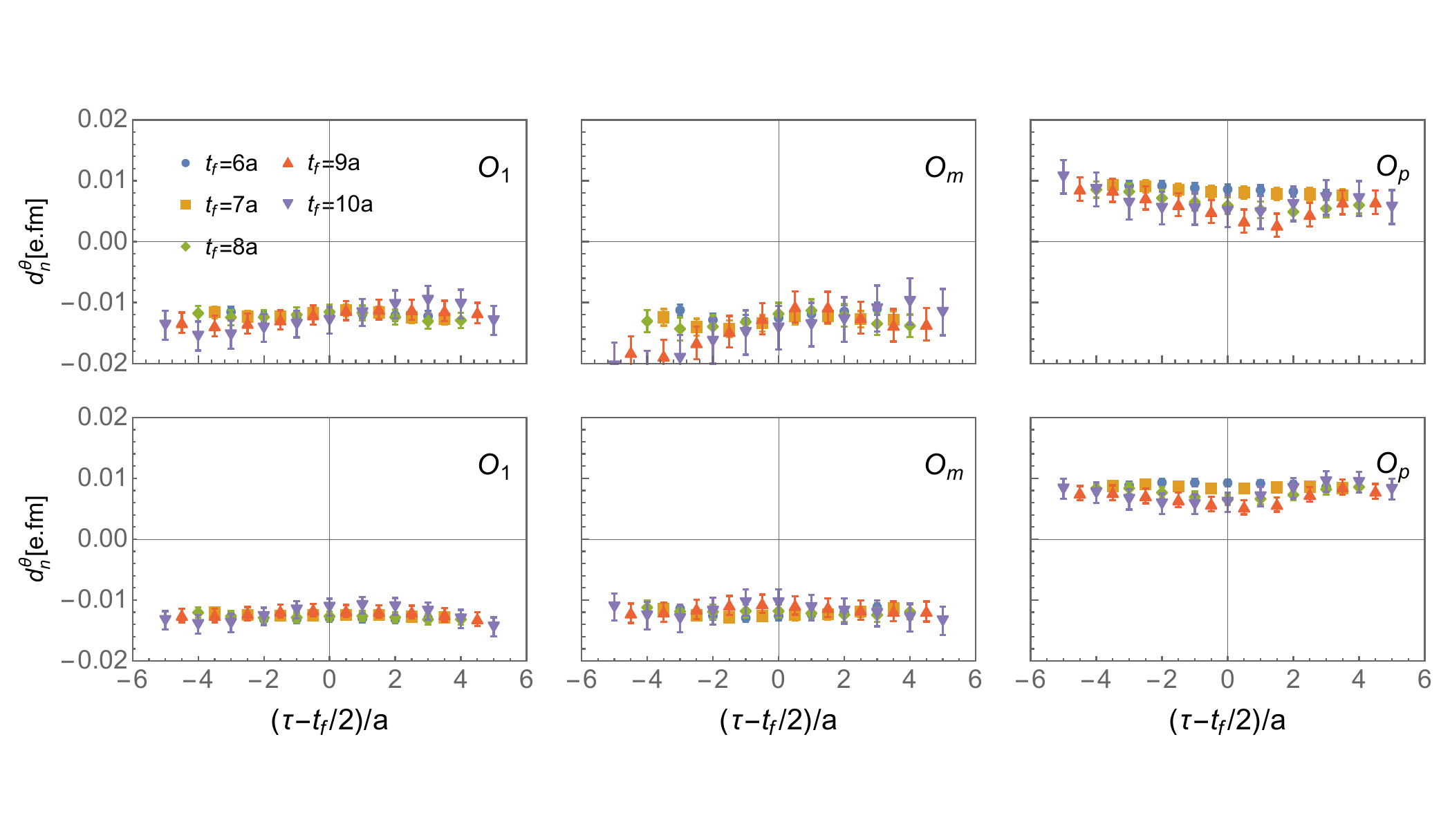}
\vspace{-12mm}
\caption{The EDM results obtained using the GEVP analysis Eq.~(\ref{eqn:edm_bgem_GEVP1}) with different interpolating
operators on ensemble 24I-005 with electric field strength $|n_z|=1$ (top) and $|n_z|=2$ (bottom).}
\label{fig:GEVPres_igf4}
\end{figure*}

The results of nEDM from the GEVP analysis are shown in Fig.~\ref{fig:GEVPres_igf4}, where the upper and lower panels correspond to different electric field strengths.
The results obtained with operators $O_1$ and $O_m$ are consistent with each other, while those
from $O_p$ have the opposite sign. Moreover, the results for $O_1$ and $O_m$
exhibit notably less excited state contamination compared to those of $O_p$.

%--------------------------------------------------------------------

\subsubsection{The energy levels of different nucleon interpolating operators}

\begin{figure*}[t]
\centering
\includegraphics[width=.96\textwidth]{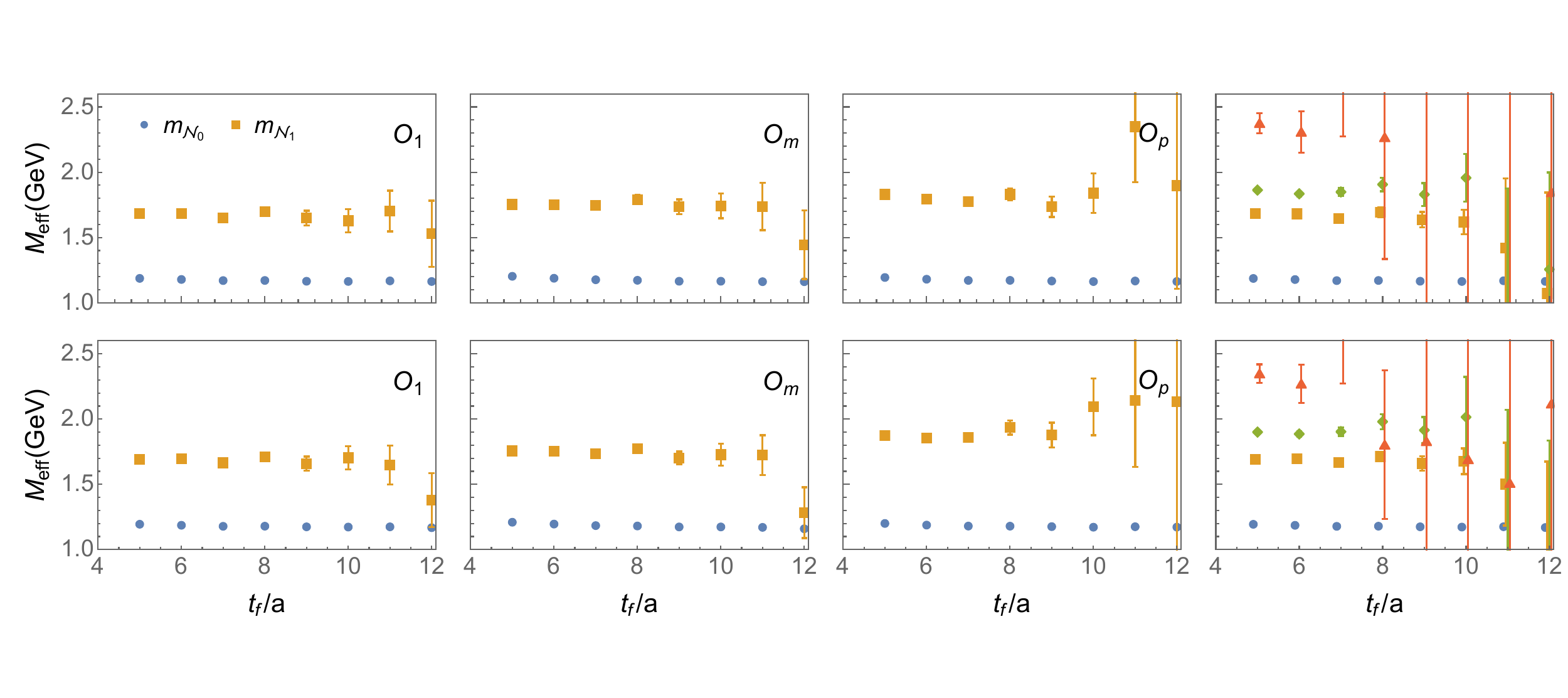}
\vspace{-12mm}
\caption{The effective masses of the ground and first excited states in the presence of electric field, obtained using
different nucleon interpolating operators on ensemble 24I-005 with electric field strength $|n_z|=1$ (top) and $|n_z|=2$
(bottom). 
The rightmost panel shows the effective masses for the
first four eigenstates obtained from the GEVP analysis using operator basis $\{O_1,O_2\}$.}
\label{fig:effmass}
\end{figure*}

In Fig.~\ref{fig:effmass}, we plot the effective masses of the ground state and first excited state in the presence of the background field,
\begin{equation}
m_{\mcN_i}(t_f)=\frac{1}{a}\log\left[\frac{\lambda_i(t_f)}{\lambda_i(t_f+a)}\right],
\end{equation}
where $\lambda_i$ is the $i$-th generalized eigenvalue obtained from the $2 \times 2$ correlator matrix of the
positive- and negative-parity components of a nucleon interpolating operator, see Eq.~(\ref{eq:2ptppbar}). 
The generalized eigenvalues behave as $\lambda_i(t_f,t_0)=e^{-m_{\mcN_i}(t_f-t_0)}$. 
These results indicate that all three interpolating operators couple to the same ground state. 
However, the mass of the first excited state for the operator $O_p$ is higher than that of $O_1$ and $O_m$. 
In the rightmost panel, we show the effective masses for the four eigenstates obtained from the four-dimensional GEVP applied to the combined basis of parity-projected operators of $\{O_1,O_2\}$.
A small mass gap between the first and second excited states is observed, while the lowest-lying excited state is not
observed in the spectrum associated with $O_p$, indicating that $O_p$ has small overlap with this state.
Therefore, a GEVP based on $O_p$ alone cannot effectively remove its contamination to the ground state matrix element.
In the following,  we explicitly examine the biorthogonal weight of different interpolating operators with these eigenstates. 

%--------------------------------------------------------------------

\subsubsection{Biorthogonal weight of different interpolating operators with eigenstates}
\begin{figure*}[t]
\centering
\includegraphics[width=.48\textwidth]{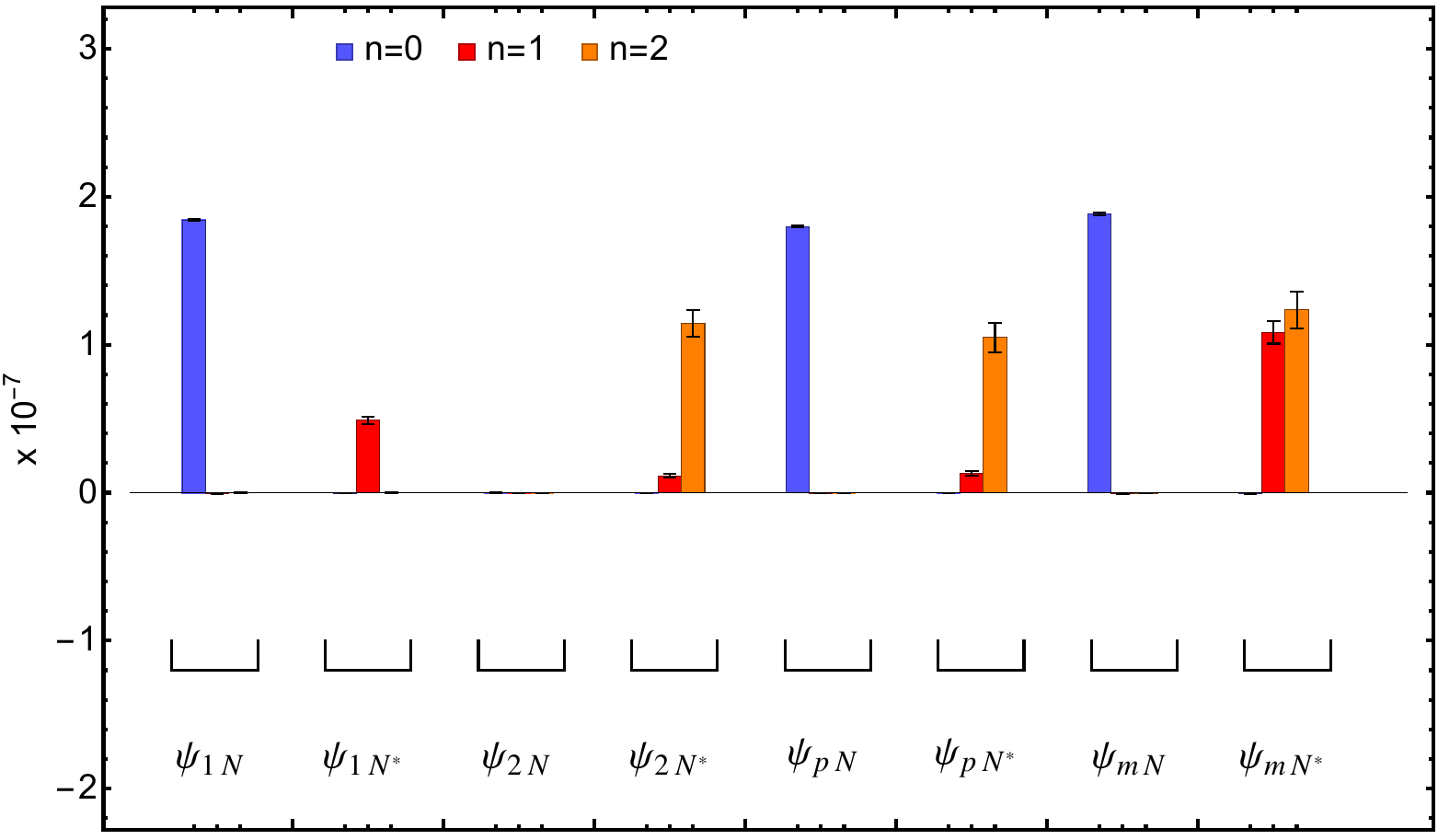}
\includegraphics[width=.48\textwidth]{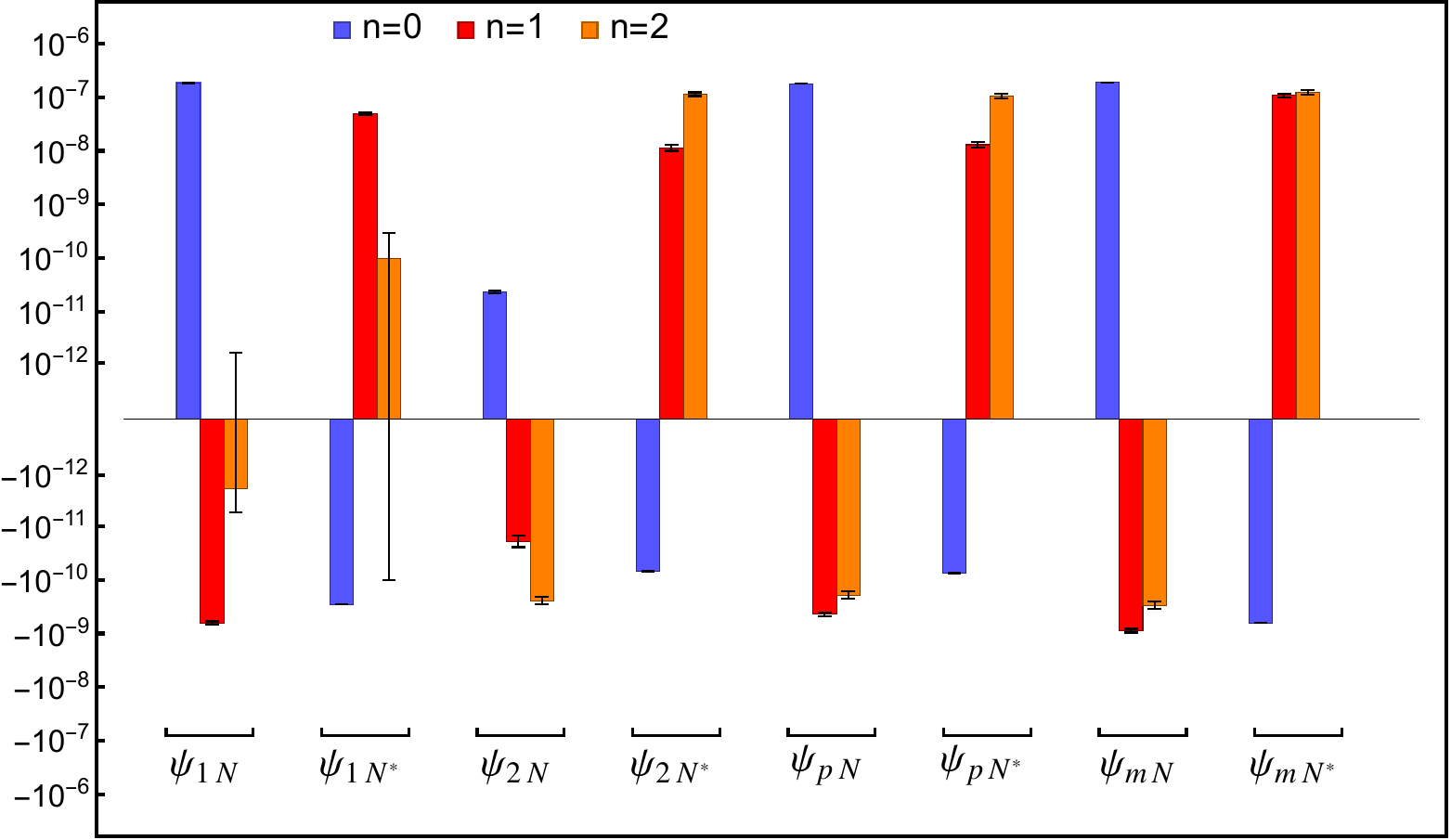}
\\
\includegraphics[width=.48\textwidth]{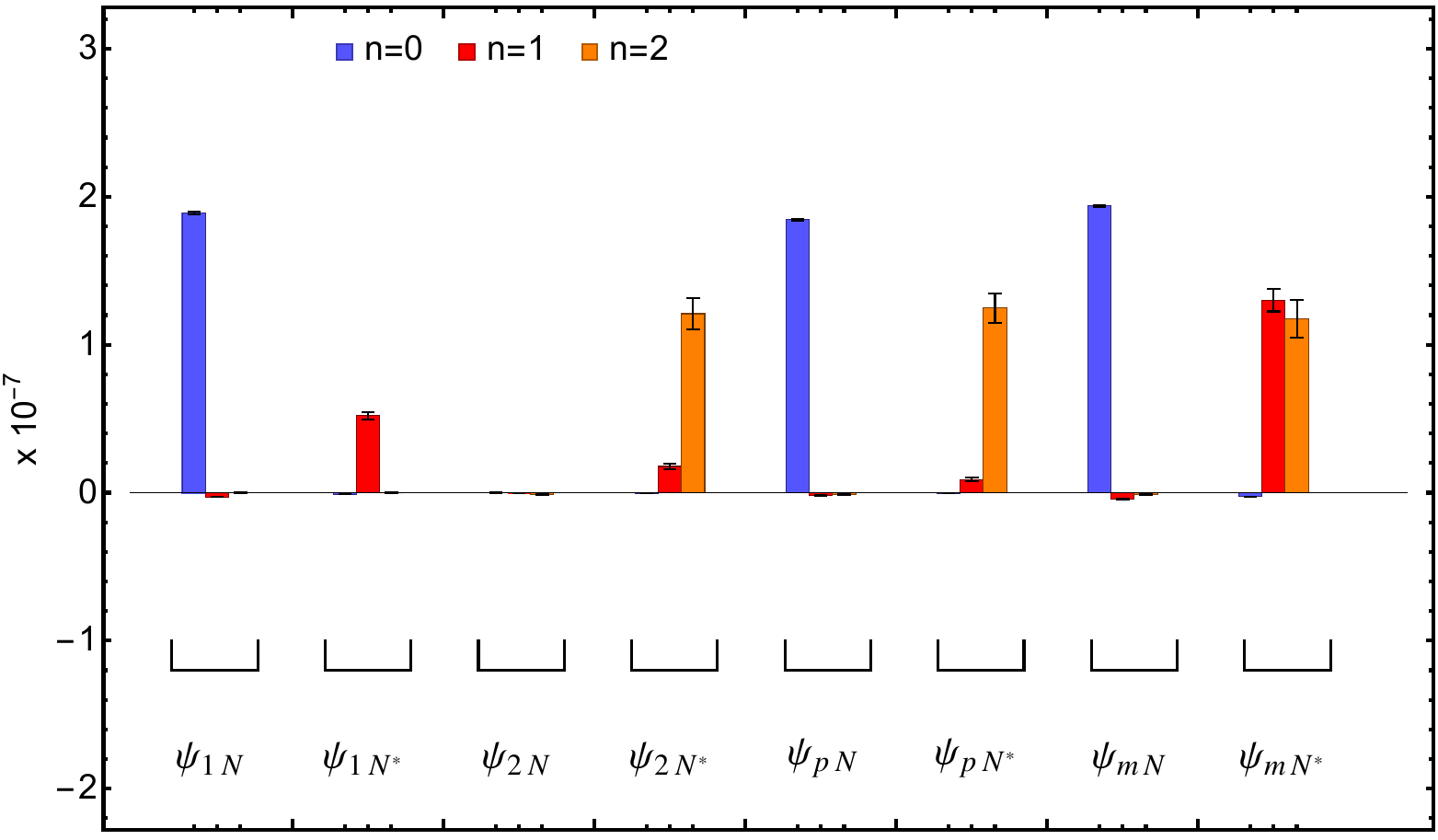}
\includegraphics[width=.48\textwidth]{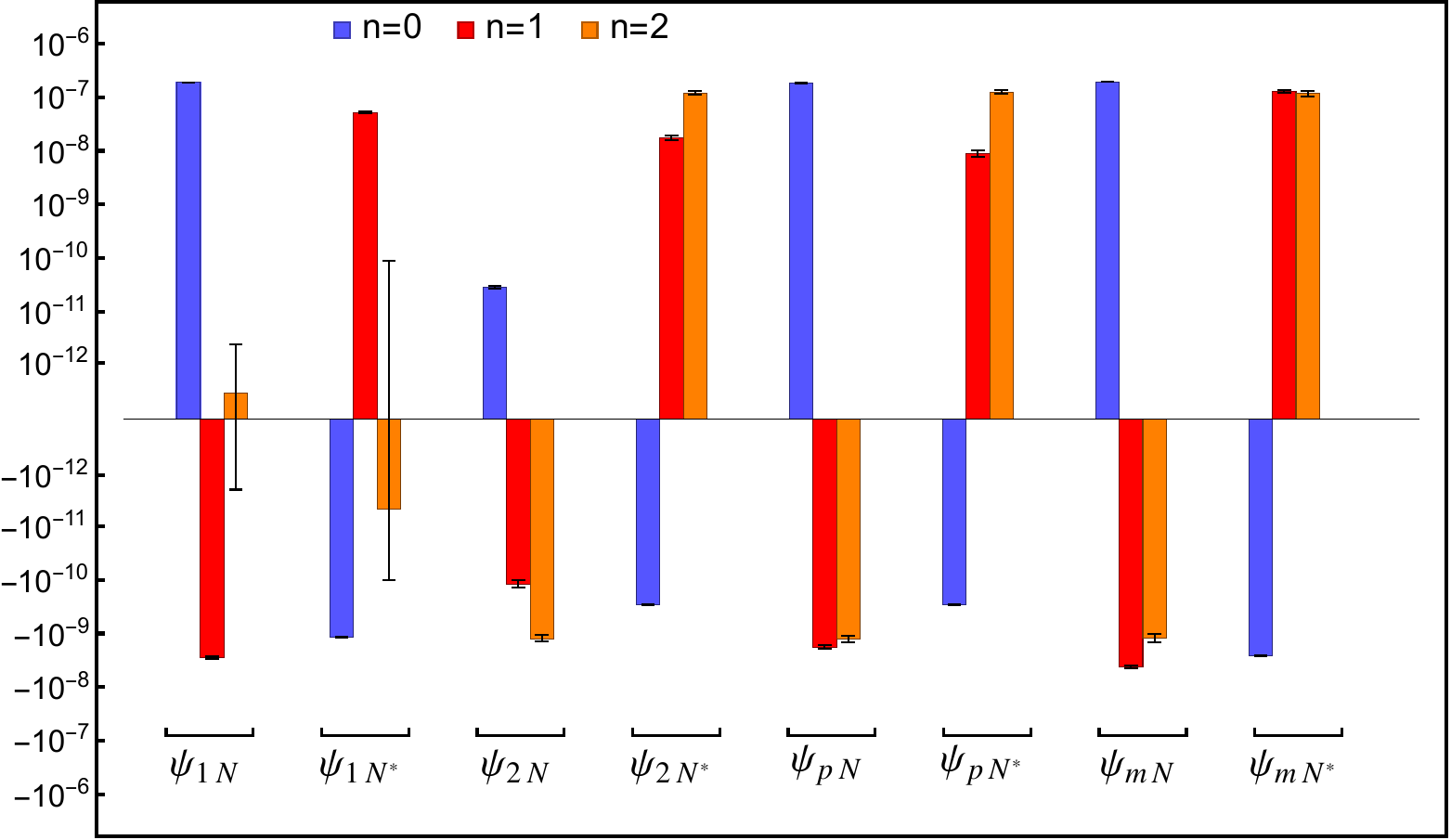}
\caption{The biorthogonal weights of the positive ($\psi_{1N},\psi_{2N},\psi_{pN},\psi_{mN}$) and negative ($\psi_{1N^*},\psi_{2N^*},\psi_{pN^*},\psi_{mN^*}$) parity components of the interpolating operators $O_1,O_2, O_p,O_m$ with the first three eigenstates ($n=0,1,2$) for electric field strengths $|n_z|=1$ (top) and $|n_z|=2$ (bottom), respectively. The plots on the right  show the same data as on the left but on a logarithmic scale.
}
\label{fig:overlap}
\end{figure*}

Once the left and right eigenvectors are obtained, they can be used to calculate the biorthogonal weight of optimal operators with the eigenstates, 
\begin{equation}
\begin{aligned}
\la \Omega^L |O^{L,\uparrow}_{\mathcal{N}_n}(0)|\mathcal{N}_n^R\rangle\langle \mathcal{N}_n^L|O^{\dagger R,\uparrow}_{\mathcal{N}_n}(0)|\Omega^R\rangle
=v^\dagger_{L,n}  C^{\psi\psi^\dag,\uparrow}_{2pt,\mcE}(t_0)v_{R,n}e^{m_{\mcN_n}t_0}
\end{aligned}
\end{equation}
To the leading order in the electric field $\mcE_z$, the polarized vacuum state can be written as
\begin{equation}\label{eq:vac}
\begin{aligned}
|\Omega^R\ra &= |\Omega\ra
+ i\mcE_z \sum_{n} \frac{\la n^-|\mcD_z|\Omega\ra}{ - E_{n}} |n^-\ra + O(\mcE_z^2) \\
\la\Omega^L| &= \la \Omega| 
+ i\mcE_z \sum_{n} \frac{\la \Omega|\mcD_z|n^-\ra}{ - E_{n}} \la n^-| + O(\mcE_z^2)\,,
\end{aligned}
\end{equation}
where $|n^-\ra$ are parity-odd states with the same quantum numbers as the dipole operator $\mcD_z$.
Since the optimized operators are defined as 
\bea
O^{L,\uparrow}_{\mathcal{N}_n}=(v^\dagger_{L})_{ni}\psi_i,~~~O^{\dagger R,\uparrow}_{\mathcal{N}_n}=\psi^\dagger_i(v_{R})_{in}.
\eea
The overlap between the original interpolator $\psi_i$ and eigenstates can be obtained from
\bea\label{eq:am2}
&&\la \Omega^L |\psi_i(0)|\mathcal{N}_n^R\rangle \langle \mathcal{N}_n^L |\psi^\dagger_j(0)|\Omega^R\rangle
\nonumber\\
&=&\sum_{k,k'}(v^\dagger_L)^{-1}_{ik}\la \Omega^L |O^{L,\uparrow}_{\mathcal{N}_k}(0)|\mathcal{N}_n^R\rangle\langle \mathcal{N}_n^L|O^{\dagger R,\uparrow}_{\mathcal{N}_{k'}}(0)|\Omega^R\rangle (v_R)^{-1}_{k'j}\nonumber\\
&\simeq&(v^\dagger_L)^{-1}_{in}\la \Omega^L |O^{L,\uparrow}_{\mathcal{N}_n}(0)|\mathcal{N}_n^R\rangle\langle \mathcal{N}_n^L|O^{\dagger R,\uparrow}_{\mathcal{N}_{n}}(0)|\Omega^R\rangle (v_R)^{-1}_{nj}
\nonumber\\
&=&(v^\dagger_L)^{-1}_{in}(v^\dagger_{L,n}  C^{\psi\psi^\dag,\uparrow}_{2pt,\mcE}(t_0)v_{R,n}e^{m_{\mcN_n}t_0})(v_R)^{-1}_{nj}
\eea
In $\{O_1,O_2\}$ GEVP analysis, the basis operators are chosen as $\psi_i=\{\psi^\uparrow_{1N},\psi^\uparrow_{1N^*},\psi^\uparrow_{2N},\psi^\uparrow_{2N^*}\}$, where $\psi^\uparrow_{iN}$, $\psi^\uparrow_{iN^*}$ denote the (spin-up) positive-and negative-parity components of the interpolating operators $O_i (i=1,2)$, respectively. The same notation also applies to the $O_p/O_m$ basis.
Using the above formula, one can calculate the biorthogonal weights of parity components of interpolating operators $O_1$, $O_2$, $O_p$ and $O_m$ with the eigenstates in the presence of the background electric field. 
The biorthogonal weights are shown in Fig.~\ref{fig:overlap}.
For example, the positive-parity component of $O_p$, $\psi^\uparrow_{pN}=\psi^\uparrow_{1N}+\psi^\uparrow_{2N}$, couples predominantly to the ground state.
Its negative-parity component
$\psi^\uparrow_{pN^*}=\psi^\uparrow_{1N^*}+\psi^\uparrow_{2N^*}$, as discussed above, couples predominantly to the second excited state ($n=2$), and its overlap with the first excited state ($n=1$) is approximately an order of magnitude smaller. 
Similarly, we also observe that the coupling of $\psi^\uparrow_{2N^*}$ to the first excited state is an order of magnitude smaller than its coupling to the second excited state. 
Furthermore, we find that the product $\la \Omega^L |\psi_{1N^*}(0)|\mathcal{N}_1^R\rangle \langle \mathcal{N}_1^L |\psi^\dagger_{2N^*}(0)|\Omega^R\rangle$ is negative, indicating that the corresponding spectral weights, $\la \Omega^L|\psi_{1N^*}|\mcN_1^R\ra$ and $\la \Omega^L|\psi_{2N^*}|\mcN_1^R\ra$, have opposite signs.
This relative sign leads to a cancellation in the combination of $\psi_{1N^*}+\psi_{2N^*}$, explaining why this component of $O_p$ has a relatively small overlap with the first excited state.

Note that some biorthogonal weights are negative as a consequence of the parity mixing effects induced by the background electric field. For example, the biorthogonal weight of the negative-parity operator $\psi_{1N^*}$ with the ground state can be expressed as
\bea\label{eq:amneg}
&&\la \Omega^L |\psi_{1N^*}(0)|\mathcal{N}_0^R\rangle \langle \mathcal{N}_0^L |\psi^\dag_{1N^*}(0)|\Omega^R\rangle
\nonumber\\
&=&-\mcE_z^2\left|\sum_i\frac{\la N_i^*|D_z|N_0\ra}{m_{N_0}-m_{N^*_i}}Z_{N^*_i}-\sum_{n} \frac{\la\Omega|\mcD_z|n^-\ra}{  E_{n}} Z^{n^-}_{N_0}\right|^2
\eea
where we have used Eq.~(\ref{eq:Npm}) and Eq.~(\ref{eq:vac}). The factors $Z_{N^*_i}$ and $Z^{n^-}_{N_0}$  are defined by $Z_{N^*_i}=\la \Omega|\psi_{1N^*}|N^*_i\rangle$ and $\quad Z^{n^-}_{N_0} = \la n^- | \psi_{1N^*} | N_0\ra\,$. It is clear that the above biorthogonal weight is negative. The numerical results also exhibit the expected $\mcE_z^2$ scaling: the magnitudes of negative biorthogonal weights obtained with $|n_z|=2$ are approximately four times those obtained with $|n_z|=1$.
%--------------------------------------------------------------------

\subsubsection{nEDM results from conventional positive parity projector}
\begin{figure*}[ht!]
\centering
\includegraphics[width=.96\textwidth]{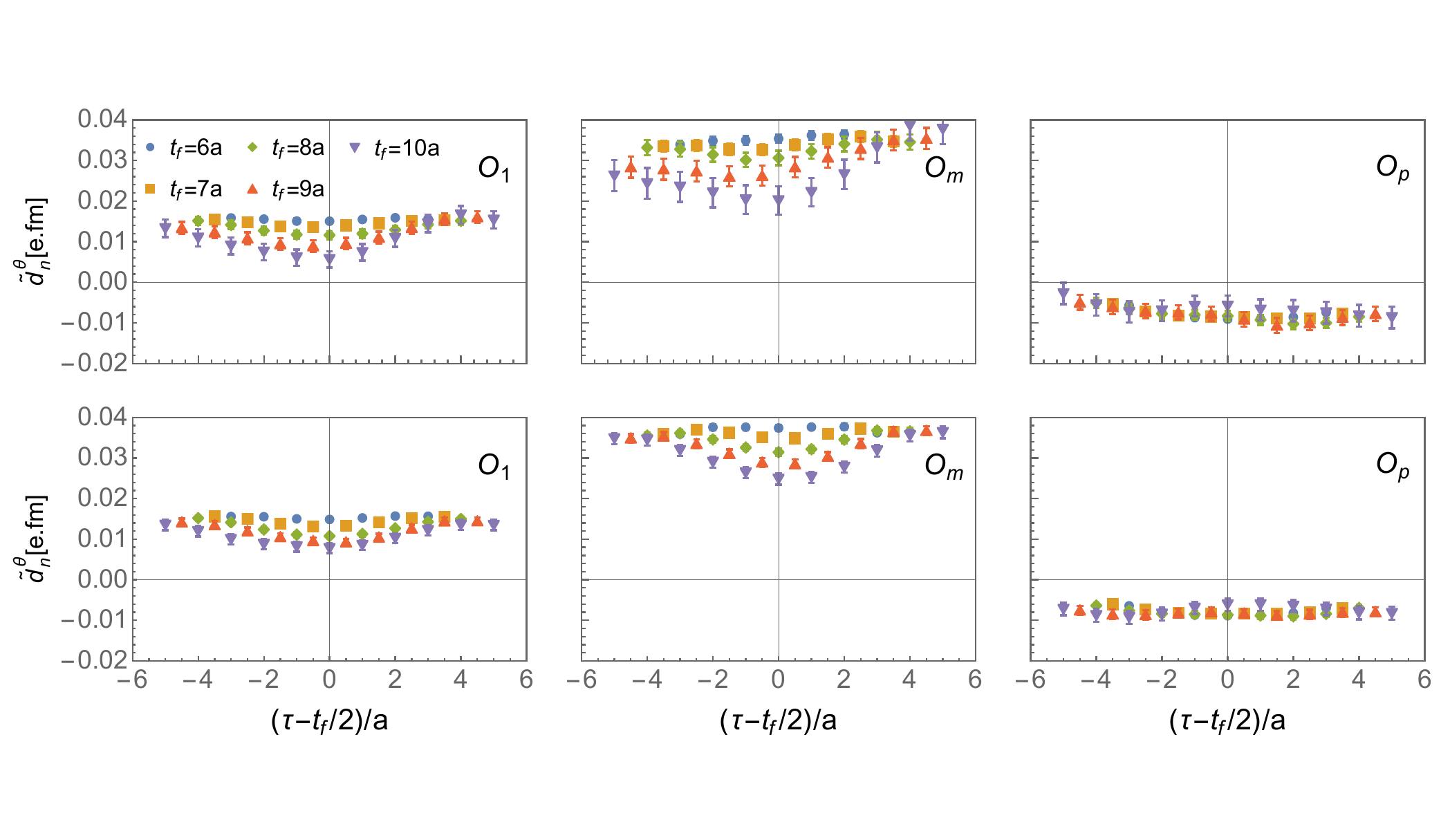}
\vspace{-12mm}
\caption{The results of $\tilde{d}_n^\theta$ defined in Eq.~(\ref{eqn:edm_tilde}) on ensemble 24I-005 with electric field strength $|n_z|=1$ (upper panel) and $|n_z|=2$  (lower panel). }
\label{fig:ori_res_igf4}
\end{figure*}

In Fig.~\ref{fig:ori_res_igf4}, we present the results using the conventional positive-parity projector applied to the different interpolating operators.
The ratio of the three-point  to two-point correlation function is defined as
\begin{equation}
\label{eqn:edm_tilde}
\tilde{d}_n^\theta(t_f,\tau)  = \frac{1}{\mcE_z} 
  \frac{\Tr \big[T^+_{S_z} \la O(t_f) \, q_{top}(\tau) \, \bar O(0)\ra_{\mcE_z}\big]}
       {\Tr \big[T^+       \la O(t_f) \, \bar O(0)\ra_{\mcE_z}\big]},
\end{equation}
where $T^+_{S_z} =T^+\cdot(1+\Sigma_z)$ with $T^+=\frac12(1+\gamma_4)$, and $O$ is the nucleon interpolating operator. 

Compared to the results from the GEVP analysis, these results exhibit significant excited state contamination, particularly for the operators $O_1$ and $O_m$. In addition, these results have the opposite sign compared to those from the GEVP analysis.
Moreover,  the results obtained from $O_1$ and $O_m$ show substantial  discrepancies, in contrast to the consistency observed in the GEVP analysis.
This behavior is expected, since parity is not a good quantum number in the presence of the electric dipole interaction;  consequently, the use
of definite-parity projectors leads to large excited state contamination. 

%--------------------------------------------------------------------

\subsection{Gradient flow time dependence}

\begin{figure}[ht!] % Fig. 7
\centering
\includegraphics[width=.4\textwidth]{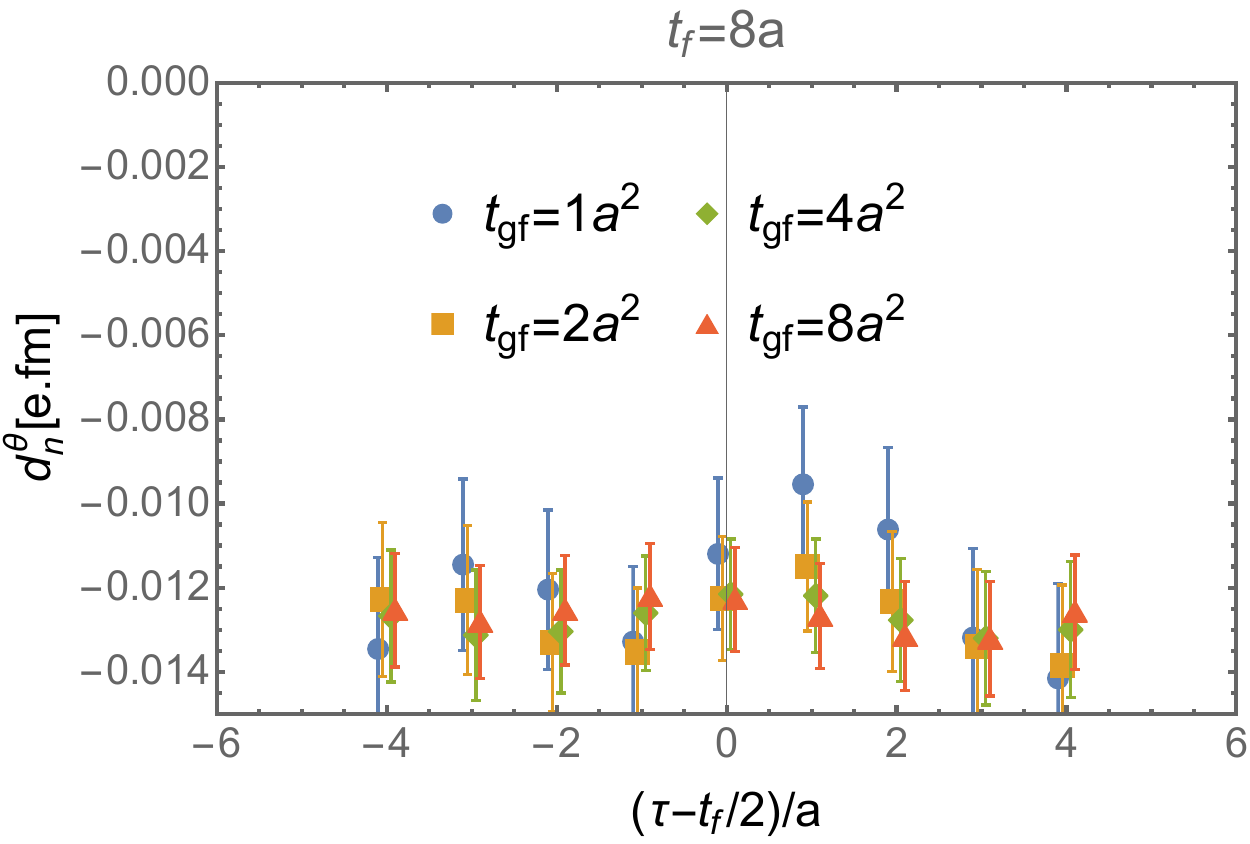}
\caption{The gradient flow time $t_{gf}$ dependence of the ratio $d_n^\theta(t_f,\tau)$ on ensemble 24I-005 with pion mass of 340 MeV,  electric field strength $|n_z|=1$, and $t_f=8a$.}
\label{fig:gf_dep}
\end{figure}

We examine the dependence of $d_n^\theta(t_f,\tau)$ on the gradient flow time in Fig.~\ref{fig:gf_dep}.
The source-sink separation is fixed at $t_f=8a$, and the results are shown for four flow times, 
$t_{gf}/a^2=1,2,4,$ and $8$. These results are obtained from a GEVP analysis 
using the full operator basis $\{O_1,O_2,O_3,O_4\}$.
Since gradient flow suppresses ultraviolet fluctuations while preserving the physical topological charge, the extracted nEDM is expected to be insensitive to $t_{gf}$ once the ultraviolet fluctuations are sufficiently tamed. 
As shown in Fig.~\ref{fig:gf_dep}, the results are indeed consistent across a range of gradient flow times. 
The data at $t_{gf}=1a^2$ exhibit larger fluctuations but remain consistent with those at larger flow times within uncertainties.
Given this stability, we adopt the results at $t_{gf}=4a^2$ as our final values.

%--------------------------------------------------------------------

\end{document}